# FLEXIBLE DATA DISSEMINATION STRATEGY FOR EFFECTIVE CACHE CONSISTENCY IN MOBILE WIRELESS COMMUNICATION NETWORKS

Kahkashan Tabassum[1] Asia Sultana[2] and Dr. A. Damodaram[3]

[1]Department of CS & IT, Maulana Azad National Urdu University, Hyd, India
`kashsss510@gmail.com`
[2]Department of IT, Muffakham Jah College of Engineering and Tech., Hyd, India
asia.sultana@mjcollege.ac.in
[3]Department of CSE, Jawaharlal Nehro Technological University, Hyd, India
`damodarama@rediffmail.com`

## ABSTRACT

*In mobile wireless communication network, caching data items at the mobile clients is important to reduce the data access delay. However, efficient cache invalidation strategies are used to ensure the consistency between the data in the cache of mobile clients and at the database server. Servers use invalidation reports (IRs) to inform the mobile clients about data item updates. This paper proposes and implements a multicast based strategy to maintain cache consistency in mobile environment using AVI as the cache invalidation scheme. The proposed algorithm is outlined as follows – To resolve a query, the mobile client searches its cache to check if its data is valid. If yes, then query is answered, otherwise the client queries the DTA (Dynamic Transmitting Agent) for latest updates and the query is answered. If DTA doesn't have the latest updates, it gets it from the server. So, the main idea here is that DTA will be multicasting updates to the clients and hence the clients need not uplink to the server individually, thus preserving the network bandwidth. The scenario of simulation is developed in Java. The results demonstrate that the traffic generated in the proposed multicast model is simplified and it also retains cache consistency when compared to the existing methods that used broadcast strategy.*

## KEYWORDS

*Network Absolute Validity Interval(AVI),Dynamic Transmitting Agent(DTA),Mobile Wireless Communication Network(MWCN)*

## 1. INTRODUCTION

The increasing demand for wireless technology and interrelated applications has encouraged companies to bring in a broad variety of wireless products such as laptops, cellular phone, etc., to serve the customer needs. Ideally, a Mobile host(MH) should be able to access the required information such as news, financial information, stock prices, etc. anytime and anywhere he wants . However, the mobile environment faces two most important restrictions, namely the system has restricted bandwidth, and the mobile user is restricted by limited resources .Hence, Caching of frequently accessed data items at the MH can become a feasible technique for providing effective service to the MH. Caching data items at the client side improves data availability in the presence of disconnection, reduces the latency in data access, relieves the bandwidth consumption, minimizes communication, saves battery life, reduces uplink requests by clients, reduces network traffic in a limited bandwidth network. Data cached at the MH should be consistent with the data at the data server that is if any modifications are done to the original copy of the data at the server database, then the local copy at MH will no longer be valid and the MH should check if the data is valid or not before answering a query. Other limitations of the mobile environment that could hamper achieving high consistency is the frequent disconnections of MH which can either be voluntarily to save battery or involuntarily due to failure or roaming. Hence, caching can be a fine technique that would enhance the system performance by decreasing the query delay. The invalidation report (IR) based approach is used to inform the users about the invalidity of the data in mobile environment .Broadcasting of IR is the usual technique for data dissemination and for maintaining cache consistency in mobile environment. This paper proposes a multicast based strategy that will disseminate data in a manner that is superior to the existing strategies. This is done by distributing AVI (Absolute validity interval) of data items to the clients and using multicasting technique for maintaining cache consistency in mobile environment. Multicasting can be an effective method which guarantees scalability, reliability and timely content distribution in wireless environment.

### 1.1. Architecture Of Mobile Wireless Communication Network (MWCN)

Mobile Wireless Communication Network (MWCN) (Fig.1) comprises of two different entities: Mobile hosts(MH's) and Fixed hosts(FH's). Few of the Fixed hosts, called the Base Stations(BS's) are wirelessly connected to the MH's. The Base Stations are connected to the server through the wired medium. Each cell in the network comprises of a Base Station to service the MH's within its cell. A MH can be moving within a cell or between cells while maintaining its network connections. There are many database servers; each database server manages one or more cells and can service only those MH's who are available in its coverage area. There can be many MH's in each cell – generating requests to get the latest copy of a data item. The servers alone can update the database. Each server broadcasts invalidation reports(IR's) periodically. Whenever a query comes in at the MH, it waits for the next IR to verify its cache contents. If data in its cache is valid, then the query is answered but if it is invalid, then a fresh copy is obtained from the server.

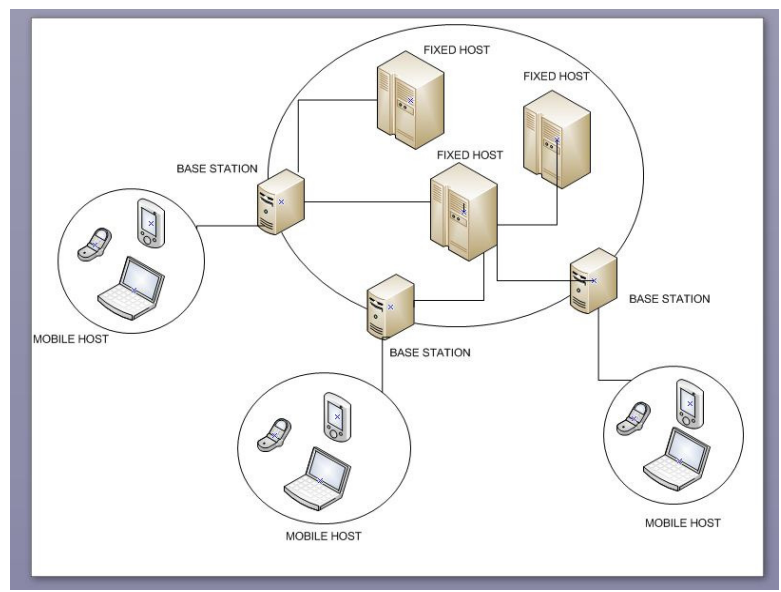

Fig. 1. Architecture of Mobile Wireless Communication Network

International Journal of Distributed and Parallel Systems (IJDPS) Vol.3, No.3, May 2012

The paper is organized as follows. In Section 2, we review related work in cache coherence. In section 3, we describe in detail our proposed system. The results of our strategy are given in Section 4, and a brief conclusion is included in Section 5.

## 2. RELATED WORK

Data items which are used frequently can be cached at the Mobile host (MH) and this technique has been recognized as a significant technique to reduce latency, traffic, communication cost in a limited bandwidth mobile environment. Majority of the studies on cache coherency for mobile environments relies on broadcast of invalidation reports (IR's) periodically. For instance, Barbara and Imielinski, proposed three variants of this approach – Broadcasting Timestamp (TS), Amnesic Terminals (AT) and Signatures (SIG) - depending on the expected duration of network disconnection [11]. However, the algorithms are efficient only if the MH's don't remain disconnected for a long period of time, duration of which is specified by the algorithm, otherwise the entire cache has to be purged even though some of the data items in the cache might still be valid. Jing et. al. proposed a bit-sequence scheme (BS) in which the IR consists of bit sequences along with a set of timestamps [13]. This approach has a disadvantage that it is too complicated and has larger IR's when compared to TS or AT methods, especially when the number of data items is larger. The problem with the above mentioned IR solutions is that they increase latency in data access, since the MH must wait for the next IR to

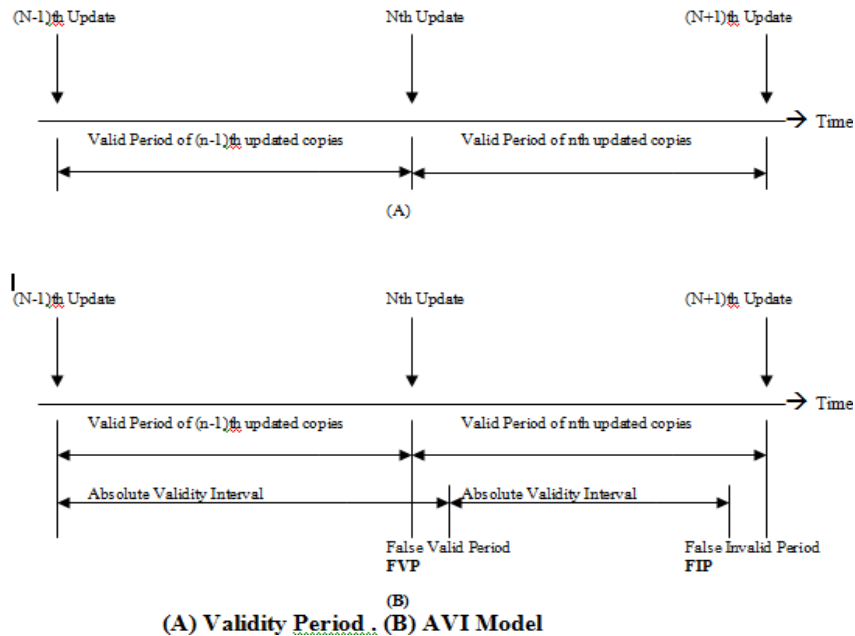

Fig. 2(A) shows Validity Period and (B) AVI Model

verify whether the data in its cache is valid or not, before answering a query. So, when a MH needs a data item which is not valid in the cache, then the MH gets the data from the server. To reduce query delay and improve bandwidth utilization ,Yuen at. al [8] proposed Invalidation by

Absolute Validity Interval (IAVI) for cache invalidation. AVI is an approximation of the life-span of the data item. Based on the intervals at which the data was updated previously, AVI is estimated. Each update has a timestamp associated with it, which specifies time at which the update took place. The time-stamp along with the AVI of the data item is used to find the validity of a data item at the MH. A cached data item, Di, at the MH is invalid if $AVI_i + TS_i <$ current time where $AVI_i$ represents AVI of a data item and $TS_i$ represents the last update time and hence new value of the data item is required. Fig. 2A explains the validity period which means that data is valid until we receive the next update that is time between two subsequent updates is the validity period of a data item. Fig. 2B differentiates between validity period and AVI. False Valid Period (FVP) is the time period where AVI overestimates the validity period of the data item and the False Invalid Period (FIP) is the time period where AVI underestimates the validity period. By suitably specifying the AVI based on the update intervals, the values of FVP and FIP can be kept to rather small values. A cached item becomes invalid when its AVI expires hence it reduces the requirement to generate explicit IR's. But to inform the MH's of the changes in AVI, IR's are generated and sent periodically. AVI approach improves cache hit probability, mean response time and reduces the number of deadline missing requests. The IAVI scheme uses IR to inform MH about the change of AVI rather than the update event of the data item and IR is sent only when AVI has reduced and not when it has increased. As a result, the size and frequency of IR can be reduced significantly. Performance studies have shown that the IAVI scheme can significantly reduce the mean response time and IR size under various system parameters.

## 3. PROPOSED SYSTEM

### 3.1. Architecture of the Proposed System

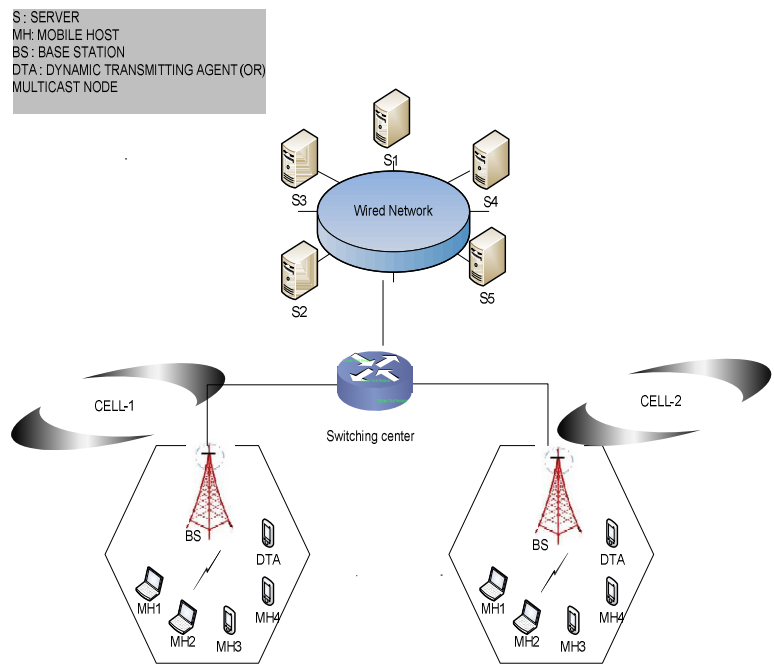

Fig. 3. Architecture of the Proposed system

The architecture of the proposed system comprises of the following entities:

1. The Mobile Host (MH)
2. The Server

International Journal of Distributed and Parallel Systems (IJDPS) Vol.3, No.3, May 20123. The Base station(BS) and
4. The DTA (selected from among the MH's based on three parameters)

## Invalidation Report

IR is an efficient method of informing the MH about the invalidity of the data when it has been updated at the server database. There are numerous methods in which IR's can be sent for achieving consistency.

In the proposed model, if the data at the server has been modified, the server informs the BS via wired links that the corresponding data is no longer valid and accordingly the BS has to identify the invalidity by broadcasting the IR, where all the MH's receiving the IR have to invalidate their corresponding data. The IR will be broadcasted only when AVI of the data item is reduced and hence size and frequency of IR is reduced.

The AVI-based approach allows for implicit invalidation of data that is the MH doesn't have to wait for IR to answer a query, it improves the mean response time and decreases the number of deadline missing requests. Our main focus is to maintain the consistency of data at the DTA. The DTA has to ask for the new version of the data from BS, which brings it from server and hands over to the DTA and DTA multicasts the updates to the MH's.

## Query Request

Whenever a query occurs, the MH searches its cache for the required data, and if a hit occurs, it verifies the validity of data by checking its AVI. If AVI + TS > current time ,then data is valid and can be returned to the user. If there is a cache miss or data is not valid then it sends a message to the BS. Upon receiving the request, the BS checks if the user has already registered with the server and the DTA. If it is registered, MH's will be getting latest updates at the multicast group address. Since DTA is a group leader ,it keeps multicasting the latest updates to the multicast group address. The data path from DTA to the MH yields a smaller delay which is our concern in this scheme.

## Roaming

Before moving to a new cell, the MH has to notify the BS about its intention of moving and its new location. The BS will maintain details of the current DTA as well as the successor DTA if at all the DTA leaves the cell.

## Query Description

1. The Mobile clients are PDA's, Laptops etc. The Base station (called server from here after) will identify one of the existing clients in the cell as the leader or DTA.
2. The factors for DTA selection are Energy level, Distance from server, Access rate.
3. The Clients send average of the values of the three factors to the server.
4. Server compares all three values sent by the clients and selects the highest among them and then broadcasts to the clients the port number of the client having the highest average value as the DTA.
5. Once the DTA is identified, that particular client requests the server for the latest updates.
6. After getting the latest updates from the server, the DTA creates a multicast socket and multicasts the data.
7. Other clients join the multicast group and receive the multicast data
8. For further queries, clients uplink only to the DTA.
9. DTA gets updates from the server which it multicasts and clients get the updates from DTA.

251

## 3.2 Flowchart

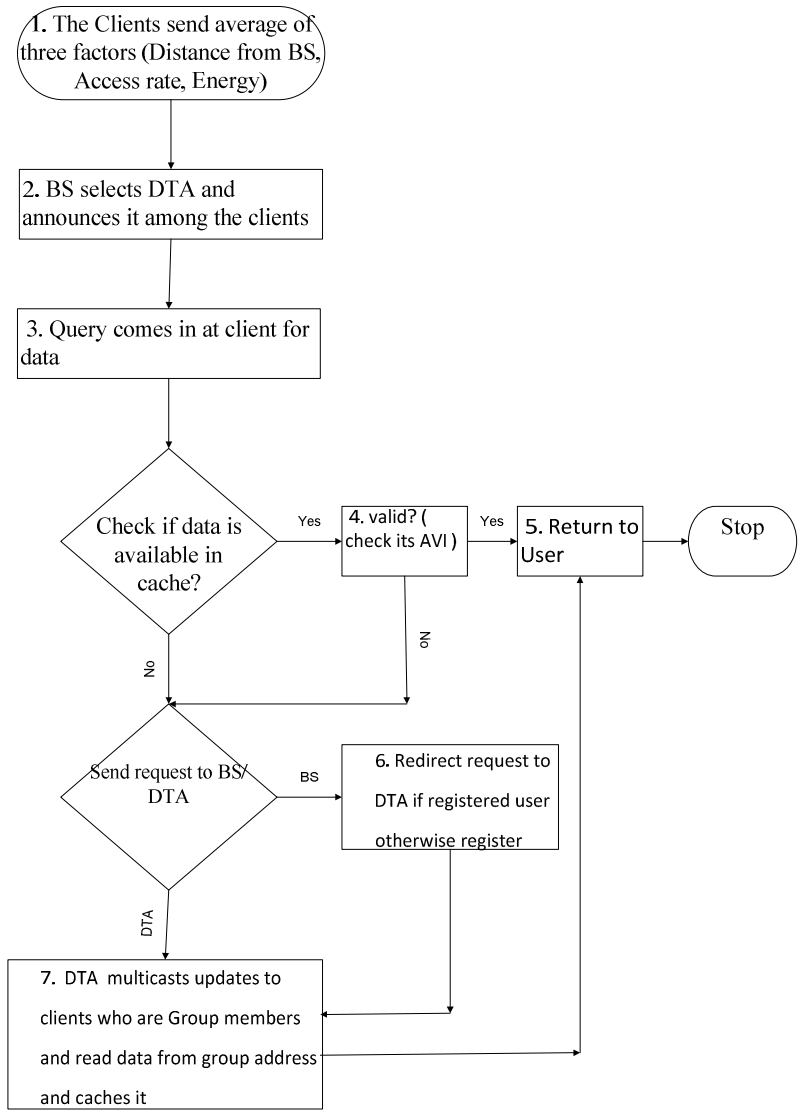

Fig 4: Flowchart depicting sequence of activities to answer a query

### 3.2.1 Base Station Algorithm

Begin
1. All the MH's send average of the three parameters namely Energy, Distance, and Access rate to the BS.
2. BS receives average of the values of the three parameters from interested MH's.
3. BS Compares all three values sent by the MH's and designates the one having the highest average value among them as the DTA.
4. BS Broadcasts the information of the DTA to all MH's within the cell.

5. BS Broadcasts IR containing information of all those data items that are updated to all MH's.
6. If MH missed Broadcast (DTA$_{nom}$, timex)
    If (BS gets query(x) from MH)
    Then BS redirects the MH to DTA.
7. BS receives the query(x) from DTA for latest updates
    If (d$_x$ is available in BS cache)
    Send Valid_data(x, dx, AVIx, t$_x$) to DTA
    Else
    Send query(x) to server.
8. If (BS gets Update(x, d'$_x$, time'$_x$) from the server)
    Update the database entry with ID x: as: d$_x$ = d'$_x$ and time$_x$ = time'$_x$

End.

### 3.2.2 Client (Mobile Host) Algorithm:

Begin
1. If (MH receives query(x) message)
    If (d$_x$ is present in MH's cache and has valid AVI)
    Answer the query.
2. If d$_x$ is not present in MH's cache with a valid AVI
    Send query(x) to BS
3. MHs register with DTA after the nomination of DTA to get specific service or updates from the DTA.
4. If d$_x$ is not present in MH's cache with a valid AVI
    Send query(x) to DTA.
5. If (d$_x$ is valid in cache of DTA)
    Receive Multicast of d$_x$ periodically from DTA
   else
    DTA Sends query(x) to BS.

End.

### 3.2.3 DTA Algorithm

Begin
1. If (DTA gets query(x) from MH)
        If (d$_x$ is valid in cache of DTA)
            Multicast d$_x$ to all clients in the multicast group
            else
            Send query(x) to BS i.e. uplink to BS on behalf of MH .
2. The MHs interested in the same information need not send query(x) message to the BS later. They obtain updates from DTA only.                                End.

## 4. RESULTS

The results are simulated using Java. The scenario comprises of three Mobile Hosts(MH's) and one Base Station (BS). One of them is selected as the DTA.

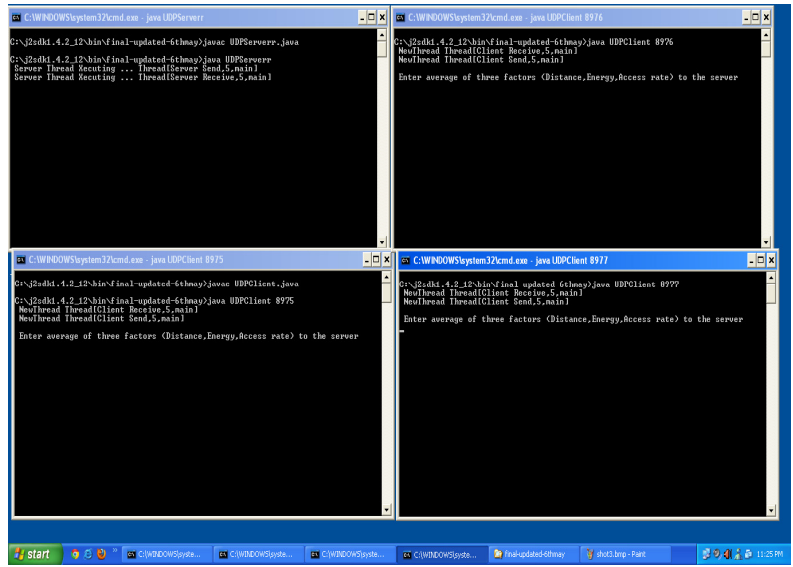

Fig. 5. Server and the 3 clients are up and running

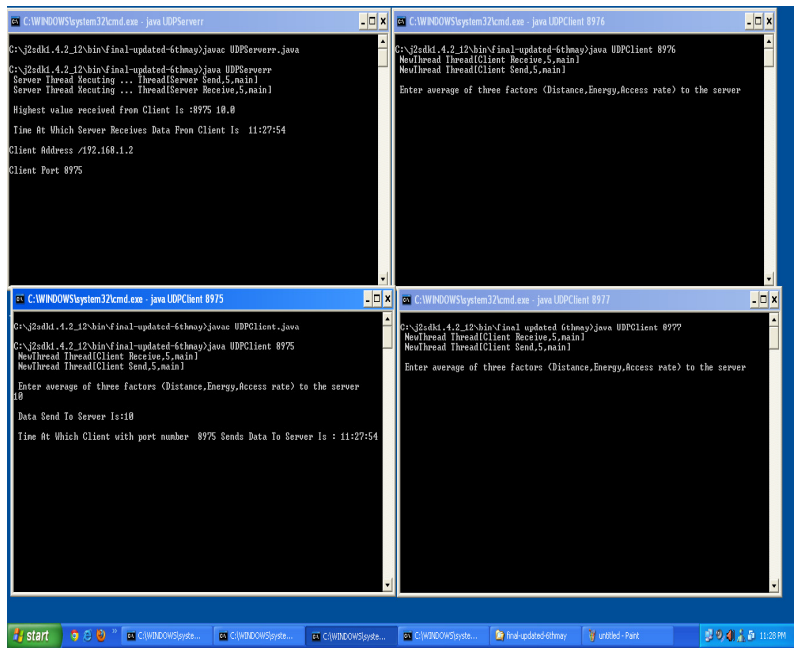

Fig. 6. Client1 sends average of three factors to the Server

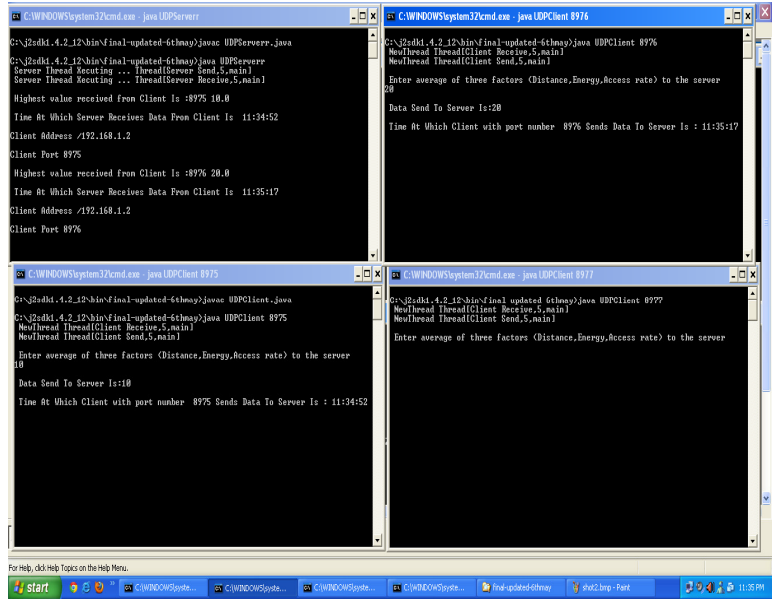

Fig. 7. Client2 sends average of 3 factors to the Server

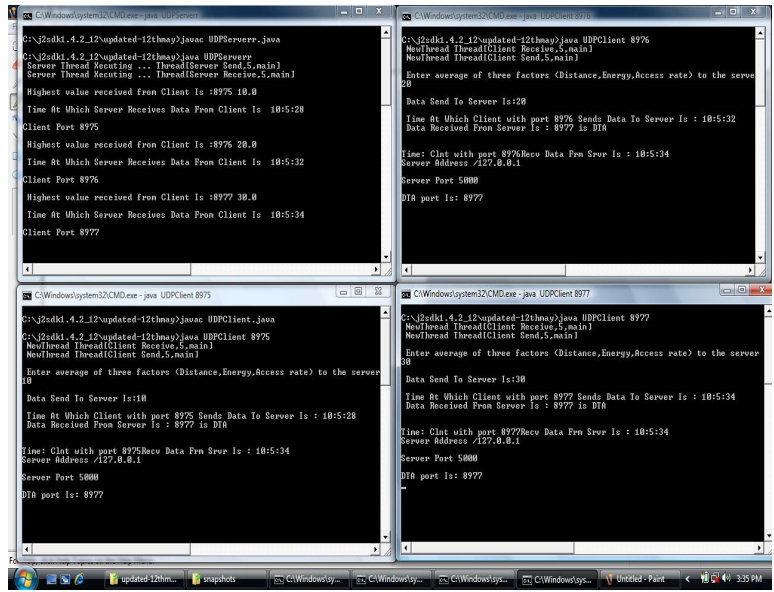

Fig. 8. Client3 sends average of 3 factors to the Server and Server selects DTA

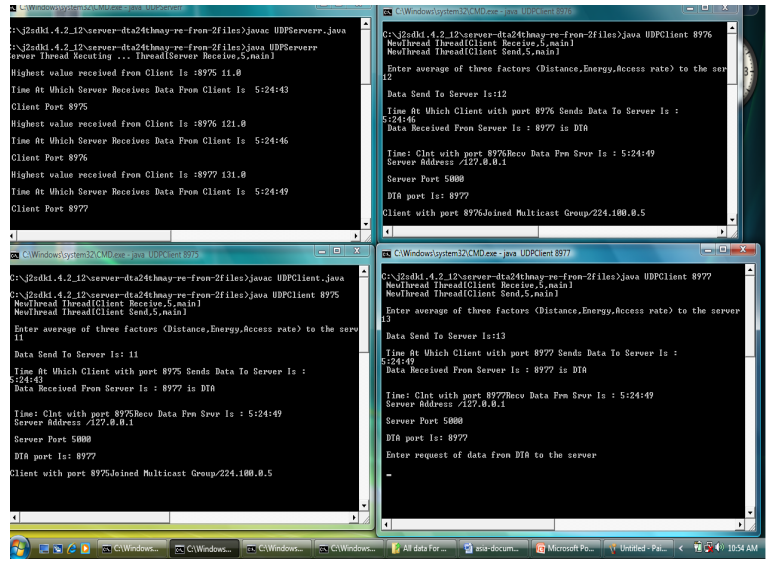

Fig. 9. DTA requests updates from Server

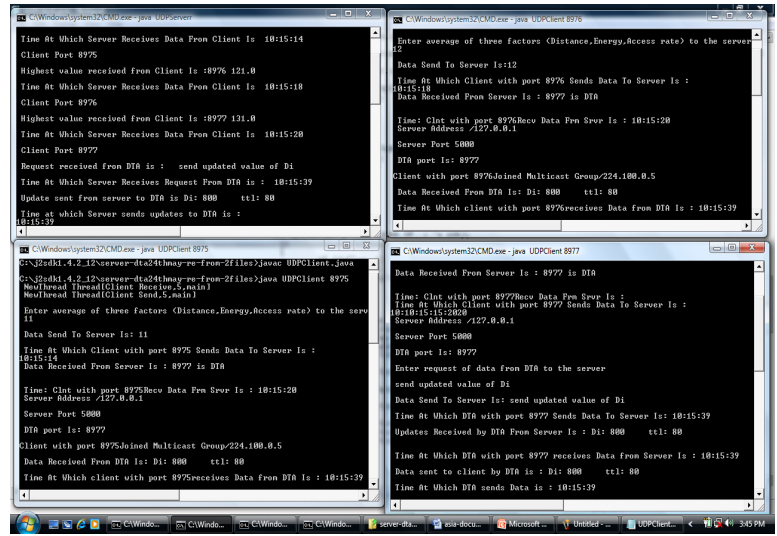

Fig. 10. The DTA gets updates from Server and multicasts to clients

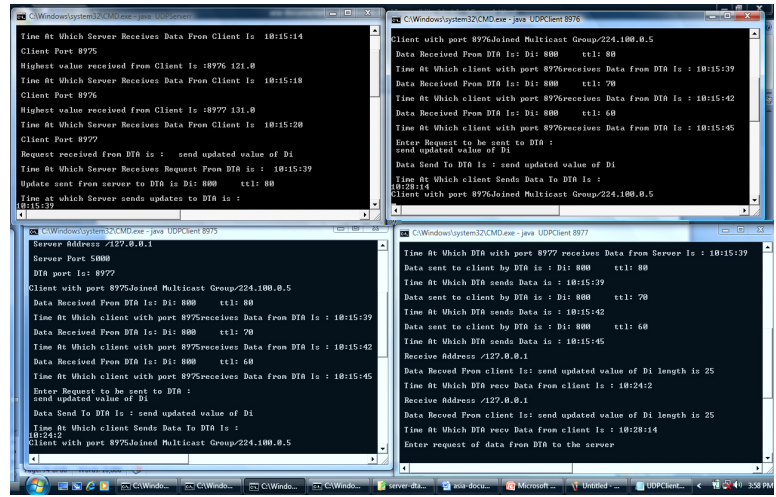

Fig. 11: Client1 uplinks to the DTA for latest updates

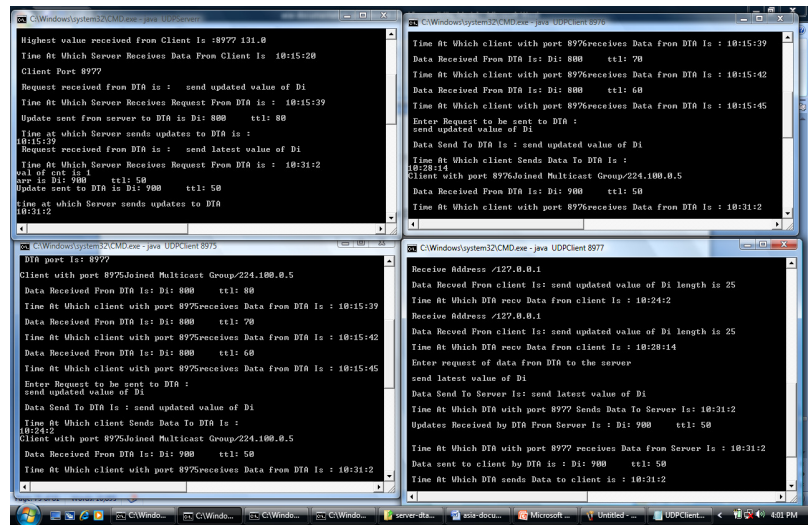

Fig. 12. Client2 uplinks to the DTA for latest updates

International Journal of Distributed and Parallel Systems (IJDPS) Vol.3, No.3, May 2012

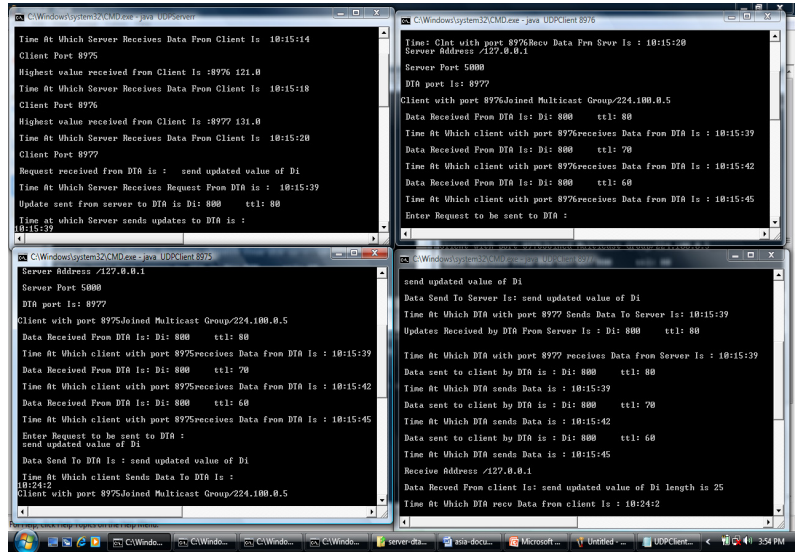

Fig. 13. DTA requests for latest updates to the Server and multicast

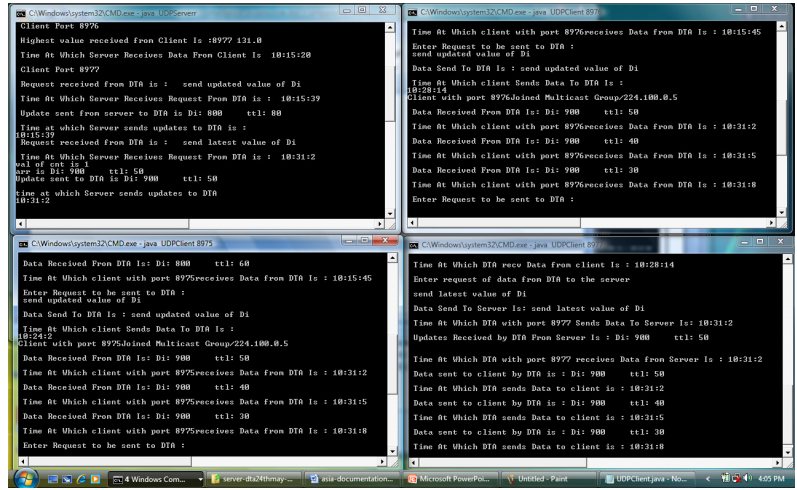

Fig. 14. DTA multicasts the updates to the clients

## 5. CONCLUSIONS AND FUTURE WORK

Cache invalidation strategies use broadcasting to distribute the information to the large population of MHs for effective cache consistency in the MWCN. Although scalable broadcasting creates lot of traffic on the network. This proposed model of Multicast data dissemination to maintain cache consistency eliminates the traffic problem and improves the availability of data. The server is relieved from the burden of servicing a group of clients because DTA can service some of the clients on behalf of server. The server can use its valuable time for other critical activities. The proposed strategy reduces the number of uplink requests to the server by the clients and also guarantees cache consistency among the data items present in clients' cache and the server.  The model implemented also minimizes the delay associated with answering query in MWCN and provides the sleeping MHs who missed the broadcast of Invalidation Reports with the latest data updates instantly. Thus all the possible resources are utilized in the best possible manner by incorporating the model discussed in this research work .

The DTA cache organization and management can be enhanced in the future. Indexing techniques can be used by the DTA to provide instant information to the clients in case the group is large. Prediction Algorithms can be used to implement cache replacement policy for DTA Cache. The methods of enhancements given above can increase cache hit ratio of DTA and thus improve the overall system performance.

Adhoc Multicast Routing can be used to route the data from DTA effectively in order to manage small and large groups. The Integration of adhoc networks to the internet and fairness in case of congestion during multicasting also requires attention for the future study that can reduce query delay to great extent.

The effectiveness of security issues imposed on DTA and energy efficient algorithms used can be open areas of research in future study.

**Authors**

**Kahkashan Tabassum** is currently working as Assistant Professor in Maulana Azad National Urdu University, Gachibowli, Hyderabad. She had worked as Associate Professor in CSED, Muffakham Jah College of Engg. and Tech., Hyderabad, India for more than 13 years. She received her Master degree in Computer Science and Engineering in the year 2005 from Jawaharlal Nehru Technological University, Hyd, India. Her area of interests are Database Management Systems, Mobile Computing, Network Security and Data Mining. She teaches B.E., M.E., M.C.A, M.B.A courses. She is Life member of CSI. She has published 22 technical papers in National, International conferences and journals.

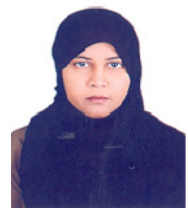

**Dr. Avula Damodaram** is the Director of School of Continuing & Distance Education, JNTU, Hyd. He joined as faculty of CSE, JNTU, Hyd in the year 1989 after completing his Ph. D from the same university. Since then he has served the university in distinguished capacities such as Professor, Head of the Department and Vice Principal. He has successfully guided many Ph.D. scholars. His areas of interest include Computer Networks and Mobile Computing. Dr. Damodaram has published 35 technical papers in National and International journals and presented 45 papers at different National and International conferences.

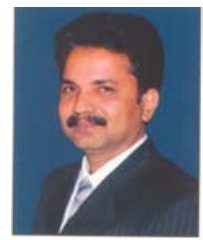

**Asia Sultana** is an Associate Professor in I.T.D in Muffakham Jah College of Engineering and Technology, Hyderabad, India. She received her Master degree in Computer Science and Engineering in 2011 from Jawaharlal Nehru Technological University, Hyderabad, India. Her area of interests include Data Structures, Database management systems, Computer Networks and Mobile Computing. She teaches and guides students at B.E level. She is life member of CSI

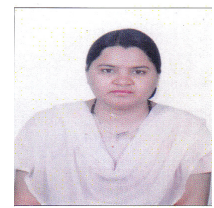